\documentclass[reprint,amsmath,amssymb,aps,pre]{revtex4-1}

\usepackage{graphicx}
\usepackage{dcolumn}
\usepackage{bm}
\usepackage{hyperref}

\begin{document}
	
\title{Optimized Finite-Time Work Protocols for the Higgs RNA-Model}

\author{Peter Werner}
\email{peter.werner@uni-oldenburg.de}
\author{Alexander K. Hartmann}%
\email{a.hartmann@uni-oldenburg.de}
\affiliation{
	Institut f\"ur Physik, Universit\"at Oldenburg, 26111 Oldenburg, Germany
}

\date{\today}
\begin{abstract}
	The Higgs RNA-Model \cite{Higgs_1996} is studied in regard to finite-time
	driving protocols with minimal-work requirement.
	In this paper, RNA sequences which at low temperature exhibits hairpins are 
	considered,  which are often cited as typical template systems in 
	stochastic thermodynamics.
	The optimized work protocols for this glassy many-particle system are 
	determined numerically using the parallel tempering method.
	The protocols show distinct jumps at the beginning and end, which have been
	observed previously already for single-particle systems
	\cite{Schmiedl_2007,Then_2008, Gomez-Marin_2008}.
	Counter intuitively, optimality seems to be achieved by staying close to
	the equilibrium unfolding transition point.
	The change of work distributions, compared to those resulting from a naive
	linear driving protocol, are discussed generally and in terms of free energy
	estimation as well as the effect of optimized protocols on rare work process
	starting conditions.
\end{abstract}

\maketitle

\section{Introduction}
Moving a system from an initial state at time $t_0$ to some final state at
$t_f$ by varying an external control parameter $\lambda(t)$ in a finite period
of time $t \in [0, \tau]$ requires an amount of work $W[\lambda(t)]$
that is in general a functional of the driving.
In classical thermodynamics for macroscopic systems, a lower bound for the
required work is given by the difference in free energy $\Delta F$ between the
equilibrium states at the initial and final control parameters
\begin{equation}\label{eq:work:free:energy:lower:bound}
W[\lambda(t)] \ge \Delta F \equiv F(\lambda(\tau)) - F(\lambda(0)),
\end{equation}
where equality holds for quasi-static, infinitely slow driving
(i.e. $\tau \rightarrow \infty$).
For mesoscopic systems, thermal fluctuations are a relevant contribution to the
path the system takes when subjected to the driving protocol $\lambda(t)$.
Hence, the work $W$ is not a simple scalar value anymore but a stochastic
quantity that fluctuates each time the work process is repeated and therefore
follows a distribution $P_{\lambda}(W)$.
Note that the distribution for small systems often have signifcant 
contributions even for $W<\Delta F$, i.e., energy might be taken from the bath.
These work distributions and their moments usually fulfill various relations,
like Crook's theorem \cite{Crooks_1999} or the Jarzynki equality \cite{Crooks_1998}
\begin{equation}\label{eq:jarzynski:equality}
\left\langle e^{-\beta W} \right\rangle = e^{-\beta \Delta F},
\end{equation}
with the inverse temperature $\beta = \frac{1}{k_\mathrm{B} T}$, 
where the scale $k_\mathrm{B}=1$ is used in the following.
The average $\langle \ldots \rangle$ is over
the initial equilibrium at $\lambda=\lambda(0)$ and all
non-equilibrium trajectories $\lambda(0)\to\lambda(\tau)$.
Applying Jensens inequality to Eq.~\eqref{eq:jarzynski:equality} gives
$\left\langle W \right\rangle \ge \Delta F$, showing that 
Eq.~\eqref{eq:work:free:energy:lower:bound} still holds merit but now in a
statistical sense.

It is worth asking whether there are \emph{optimal} protocols which meet the lower bound on the work $W$ given by the equilibrium free energy difference 
$\Delta F$, and, if they exist, how they look like.
For once this is useful because they can improve the estimation of said
free energy difference via Eq.~\eqref{eq:jarzynski:equality} \cite{Geiger_2010}.
Also, they could be an alternative or supplementary approach to
large-deviation methods, e.g. optimizing the protocol instead of employing a
more complicated large-deviation algorithm \cite{Hartmann_2014, Werner_2021}.
The design of biological nanoscale machines is another aspect, where evolution
might have designed them to be energy efficient and hence follow an
optimized protocol.

With respect to optimal protocols, in past works mostly systems with few 
degrees of freedom, like single one-dimensional colloidal
particles, have been considered so far \cite{Schmiedl_2007, Then_2008,
Gomez-Marin_2008, Solon_2018, Xu_2021, Blaber_2021}.
Some of the work includes numerical optimization, e.g.,
using threshold accepting \cite{dueck1990} applied to a Brownian particle 
\cite{Then_2008}, or genetic algorithms \cite{goldberg1989, michalewicz1994} 
applied to cyclic protocols for microscopic heat engines\cite{Xu_2021}.
With respect to controlling systems, which exhibit many degrees of freedom and 
collective behavior, the Ising-model has also been studied, with temperature as 
an additional control parameter and the aim to invert the magnetization of a
given system.
Where notable contributions are restricted to the linear response regime
\cite{Rotskoff_2015}, or are in the context of parametrized protocol ensembles 
\cite{Gingrich_2016}.

With respect to the shape of optimal protocols, discontinuous jumps of
the  control-parameter value at the beginning and end are a 
reoccurring observation.
Still, even moderate changes to simple systems can result in new and surprising 
features in the space of optimized protocols, e.g.: delta peaks
for the underdamped langevin particle in a harmonic trap 
\cite{Gomez-Marin_2008},
regions with single or multiple jumps for a dipole potential \cite{Then_2008},
and the existence of phase transitions between optimal-work and
optimal-work-fluctuations protocols \cite{Solon_2018}.

In the present work the Higgs RNA-model\cite{Higgs_1996} is studied with 
numerical simulations \cite{practical_guide2015}.
To the authors knowledge, it has not been investigated with regard to
optimal protocols, despite its computational accessibility \cite{Nussinov_1978,nussinov1980,Mueller_2002}, see below,
its many degrees of freedom, and its rich 
behavior like the occurrence of phase transitions \cite{bundschuh2002,Mueller_2002,stacks2005} or 
finite-size glassy phases \cite{pagnani2000,comment_RNA2001}.

Further, it is conceptually close to experimental setups of single DNA hairpin
experiments that try to implement optimized protocols \cite{Tafoya_2019}.
This makes the model, beyond its importance for molecular biology,
an ideal playground to investigate various fundamental questions 
concerning optimized protocols, e.g.:
How are specific system state trajectories affected?
What role do phase transitions play when it comes to optimality of protocols?
Are there general rules on how work distributions change, especially in the
large-deviation regime?
How do optimized protocols change with regard to the energy landscape,
which is determined by the primary structure of the RNA molecule?
Given the space and time constrains, this paper addresses only  
some of these questions but it is clear that the RNA model is useful also
for further investigations beyond this proof of principle study.

The structure of this paper is as follows:
First, the Higgs RNA-model is described, followed by the algorithms used
to implement the work process as controlled by a given work protocol.
Next, the parallel tempering scheme is described, used to find optimized 
protocols.
In the main part, findings are presented, which include the shape of 
the optimized  protocols, mean trajectories of system observables and resulting 
work distributions.

\section{Model}
The \textit{primary structure} of a RNA molecule is given by a linear sequence 
$\mathcal{R}=(r_i)_{i=1,\dots,L}$ of $L$ bases
$r_i \in \left\lbrace \mathrm{A,C,G,U} \right\rbrace $, where the four letters 
correspond to the bases Adenine, Cytosine, Guanine and Uracil. 
Two bases $r_i$, $r_j$ at positions $i$ and $j$ 
(with $1 \le i < j \le L$) can pair to each other with
energy $e(r_i,r_j) < 0$, where the set of all pairs $(i,j)$ 
then forms the so called
\textit{secondary structure} $\mathcal{S}$.
It has four constraints:\\
(1) A base can pair to at most one other base.\\
(2) Only \emph{complementary} (Watson-Crick) A-U and C-G pairs are allowed.\\
(3) \textit{Pseudoknots} are excluded, i.e., two arbitrary pairs
$(i, j), (i', j') \in \mathcal{S}$ with $i<i'$ must either fulfill 
$i<j<i'<j'$ or $i<i'<j'<j$.\\
(4) There is a minimum distance $|j-i| > s$ between two paired bases
(here $s=2$).
\\
For bases $r,r'$, the pairing energy is set to $e(r,r') =-1$ when 
$r$ and $r'$ are complementary and  $e(r,r') =+\infty$ otherwise.
To allow for manipulation of the system, the protocol
$\lambda(t)$, which takes the role of an external force parameter, couples to
the \textit{free length} $n=n(\mathcal{S})$ of the secondary structure
$\mathcal{S}$, yielding an additional  energy of $-\lambda(t) \times n$.
The free length $n(\mathcal{S})$ is defined as the number of bases with an
sequence index that does not lie between those of any base-pair currently a part
of the secondary structure $\mathcal{S}$:
\begin{align}\label{eq:free:length}
	n(\mathcal{S}) &= \sum_{i=1}^{L} c_i\\
	c_i &= \left\lbrace
	\begin{array}{cl}
		0 & \textrm{if} \  \exists (j,k)\in \mathcal{S}\, :\, j<i<k\\
		1 & \textrm{else} \\
	\end{array} \right. \nonumber .
\end{align}
The total energy is finally given by the contributions from all base pairs and
the external manipulation:
\begin{equation}\label{eq:total:energy}
	E(\mathcal{S},\lambda) = \sum_{(i,j)\in \mathcal{S}} e(r_i,r_j)
	- n(\mathcal{S}) \lambda,
\end{equation}
where the explicit dependence on the primary structure is omitted.
This model has been used previously in other statistical mechanics frameworks
\cite{Mueller_2002,Werner_2021} and is a strong simplification compared to those
when describing real RNA. 
Dedicated software packages like ViennaRNA \cite{ViennaRNA_Softwarepackage}
exist that incorporate interactions of natural RNA more comprehensively.
But using such a sophisticated model would add details related to RNA's 
biological function wich is outside the scope of the more fundamental 
statistical-mechanics questions that are of interest here.

In the following, secondary structures in equilibrium are drawn from a
canonical ensemble at temperature $T$, i.e., they occur with the Boltzmann
probability $P(\mathcal{S})\sim \exp(-E(\mathcal{S},\lambda)/T)$.
In equilibrium, typical RNA sequences exhibit an folding-unfolding transition
at a sequence and slightly temperature-dependent critical value $\lambda_c(T)$.

\section{Algorithms}
\subsection{RNA work process}
The work process is realized as a non-equilibrium Monte-Carlo simulation,
where the duration $\tau$ of the process is proportional to the total number
of MC-Sweeps $n_\textrm{MC}$, i.e. $\tau \sim n_\textrm{MC}$.
One MC-step is the insertion or removal of a single base-pair in the
secondary structure $\mathcal{S}$ and one MC-sweep consists of $L/2$ MC-steps.

A major advantage, at this point, of the described RNA-model is that the
equilibrium behaviors, i.e., the
canonical partition function is calculable via dynamic programming in
$\mathcal{O}(L^3)$.
With this comes the ability to efficiently sample the initial equilibrium states
following the Gibbs-Boltzmann distribution directly in $\mathcal{O}(L^2)$
\cite{Nussinov_1978, Werner_2021}.
This allows to start every work process with a secondary structure sampled
in equilibrium.
For details on the sampling see appendix
\ref{appendix:equilibrium:sampling:free:length}.

Given a protocol $\lambda_j = \lambda(t_j)$ at discrete points in time
$t_j=j \Delta \tau$ with $j=0,\dots, n_{\tau}$, $\Delta \tau=\tau/n_{\tau}$, 
the work process then goes as displayed in Fig.~\ref{fig:algorithm}.

\begin{figure}[ht]
\begin{minipage}{\textwidth}
  \begin{tabbing}
    \hspace{0.2cm} \= \hspace{0.2cm} \= \hspace{0.2cm} \= \\
    \textbf{algorithm} $W\left[ \lambda(t)) \right] $\\
    \textbf{begin}\\
    \> draw for $\mathcal{R}$ an equilibrium structure $\mathcal{S}$  at \\
    \>\>   initial protocol value $\lambda_0$ and RNA temperature $T$ \\
    \> $W=0$ \\
    \> \textbf{for} $j=1,\cdots,n_{\tau}$\\
    \> \textbf{begin}\\
    \> \> $\Delta \lambda=\lambda_{j}-\lambda_{j-1}$\\
    \> \> $W = W -n(\mathcal{S})\Delta \lambda$\\
    \> \> perform $L n_{\rm MC}/(2n_{\tau})$ MC-steps over $\mathcal{S}$\\
    \> \textbf{end}\\
    \> \textbf{return}($W$) \\
    \textbf{end}	
  \end{tabbing}
\end{minipage}
\caption{Algorithm to perform an unfolding process resulting in a
work $W$. 
\label{fig:algorithm}}
\end{figure}

Note that the work process allows for explicit incorporation of jumps
at, e.g., the beginning and end of the protocol, since these are believed to be 
a generic feature of optimal protocols \cite{Schmiedl_2007}.
This was done in other numerical studies on the topic, too \cite{Solon_2018}.
For comparison, also a naive linear protocols is studied with
$\Delta \lambda = (\lambda_\tau - \lambda_0) / n_\tau = \mathrm{const.}$
as presented below. 

\subsection{Optimization Algorithm}
Following \cite{Then_2008}, the work protocol $\lambda(t)$ is discretized by
$n=42$ control points that are varied by the optimization scheme to 
approximate the optimal protocol.
For a representation of the protocol at finer resolution these $n$ points are 
interpolated linearly in between, except for the possible explicit
jumps at start and end, to yield a total of $n_{\tau}=200 + 2$ points, including
the fixed start $\lambda_0=0$ and end which is chosen to be $\lambda_{\tau}=2$.

The objective function of the optimization process for the
protocol $\lambda(t)$ is the average work
$\overline{W} = \frac{1}{N} \sum_{i=1}^{N} W_i$, determined by a finite 
number $N$ of work process executions, yielding $N$ work values $W_i$.
\\
The optimization occurs in two stages.
For both stages, Markov-chain Monte Carlo simulations with the
Metropolis-Hasting algorithm are performed in the space of protocols.
For any given protocol, exhibiting mean work $\overline{W}$,
trial protocols are constructed by randomly selecting one of
the $n$ variable protocol points $\lambda_{j}$ and changing it by a random
amount according to $\lambda_{j}\leftarrow |\lambda_{j} +
2\varepsilon \delta \lambda | \pmod {2\lambda_\tau}$, which are periodic
boundary conditons for $\lambda_{j}$ that could assist to find ceratin features,
especially delta peakes, when sampling the protocol space.
Where $\varepsilon \in (0,1]$ is a uniformly distributed random number and
$\delta \lambda$ is a magnitude chosen as described below.
The final optimization result is not expected to change when alternatively 
using more commonly employed reflective boundary conditions, i.e., moves are 
rejected when  moving outside the allowed interval.

For the trial protocol, $N$ work processes are performed
and a corresponding mean work $\overline{W}'$ is obtained.
The trial protocols are accepted for the Markov chain with an 
Metropolis probability $\min\{1, \exp[-(\overline{W}'-\overline{W})/\theta]\}$, 
which corresponds to a Boltzmann distribution of the mean work values for a given artificial ``temperature'' $\theta$.

For the first stage, the parallel tempering method \cite{Earl_2005} at 100 
artificial temperatures $\theta_{i}$ with logistic spacing is used, i.e., 
$\theta_i=\theta_0 (\delta\theta)^i$, ranging from $\theta_0 = 8 \cdot 10^{-4}$
to $\theta_{99} =50$.
The control parameter $\delta \lambda$ is chosen for each value  $\theta_i$  
such that the empirical acceptance rate is around 0.5 for this move,
respectively.
Here, $N=1000$ independent work processes  are generated to calculate the
mean work of every protocol.
The tempering is run until all replicas could visit a sufficient fraction of
all temperature sites.

The second stage is given by an annealing simulation at $\theta=0$ with an
increased number $N=10^5$ of work processes, starting from the best protocol
found in the previous stage.
For $\delta\lambda$, the same value is used as it was at $\theta_0$ in the 
first stage.

One caveat with this approach, in order to work, is to fix the random numbers
used to generate the work processes, i.e. resetting the random number generator
to the same initial state before starting sampling, which is conceptually
similar to what is done in \cite{Then_2008} with the noise history.
This is necessary, because two independent evaluations of the mean work,
performed for the same work protocol, must yield exactly the same value of 
$\overline{W}$

The numerical optimization does not necessarily yield an exact optimal protocol
due to the finite work sample size and the random nature of the algorithm
itself, i.e. maybe the optimum gets never selected within protocol space.
But even close to optimal protocols, as discussed in \cite{Gingrich_2016}, will
likely poses the distinct features, e.g. jumps, that are responsible for
achieving optimality.

\section{Results}
Three primary structure sequences of length $L=40$ are investigated
(see Tab.~\ref{tab:primary:structure:sequences}). Note that calculation
of partitions functions and sampling can be achieved for much larger
system length $L$, but here we optimize over the protocols, such that every
single optimization step consists of a full simulation according to
the algorithm shown in Fig.~\ref{fig:algorithm}. Still, the system
investigated here consist of several interacting particles, in contrast
to most work in the literature.

\begin{table}
	\caption{\label{tab:primary:structure:sequences}Overview of investigated
	primary structure sequences $\mathcal{R}$}
	\begin{ruledtabular}
	\begin{tabular}{cc}
		\emph{hairpin} & $(\mathrm{AC})^{10} (\mathrm{GU})^{10} =
		\mathrm{ACAC\dots GUGU}$ \\  
		\emph{continuous} & $(\mathrm{ACGU})^ {10}$ \\  
		\emph{asymmetric} & $(\mathrm{AC})^{5} (\mathrm{GU})^{5} (\mathrm{ACGU})^{5}$\\ 
	\end{tabular} 	
	\end{ruledtabular} 
\end{table}
The first considered sequence can fold into a perfect \textit{hairpin} secondary structure at low temperature $T$ and small force $\lambda$.
The second sequence is constructed by repeating the four bases over and over
again, i.e., it is a \textit{continuous} sequence.
The third sequence represents a combination of the \emph{hairpin} and 
\emph{continuous} sequence, which is called \textit{asymmetric} here.
Sequences like these, which all can fold into secondary structures with one ore 
more hairpins, are relevant due their abundant occurrence in nature and 
treatment in previous studies \cite{Bundschuh_1999, Gerland_2001, Faber_2013,	Liu_2006, Liu_2005, Liphardt_2002, Collin_2005, Liphardt_2001}. 
The work simulations are performed at two temperatures
$T \in \left\lbrace 0.3, 1.0 \right\rbrace$ which are representative
for the sequences being in or close to the ground state, or
where several relevant secondary structures contribute, respectively.
The three considere sequences exhibit folding-unfolding
transitions at critical values near $\lambda_c=0.5$ for $T=0.3$  
and near $\lambda_c=0.7$ for $T=1$.

\subsection{Optimized Protocols}
\begin{figure}
	\includegraphics[width=0.49\linewidth]{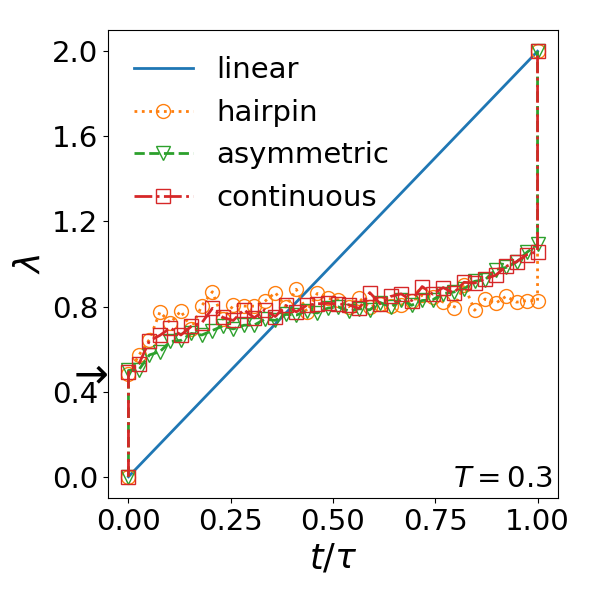}
	\includegraphics[width=0.49\linewidth]{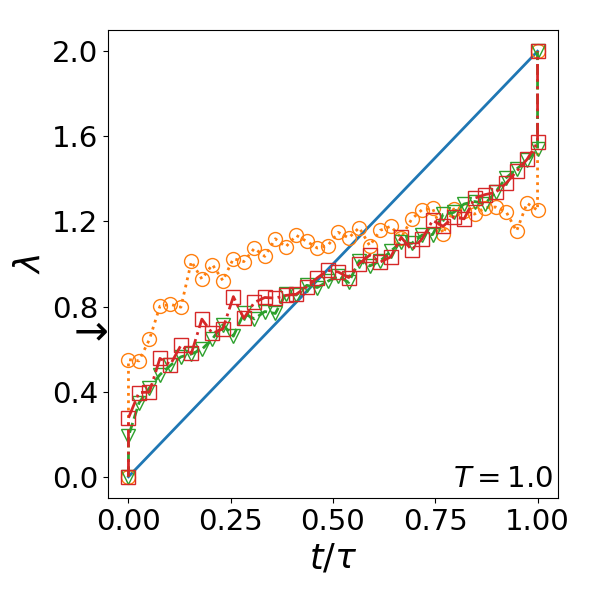}
	\caption{\label{fig:optimized:protocols:force:time}Optimized protocols
	$\lambda$ as a function of time $t$, which display distinct jumps at
	the beginning and end. The linear protocol is also shown for reference.
	The arrow on the vertical axis indicates the critical value $\lambda_c$ of the folding-unfolding transition .
	Left: $T=0.3$. Right: $T=1.0$.}
\end{figure}

Fig.~\ref{fig:optimized:protocols:force:time} shows the optimized
protocols as a function of time and the naive linear protocol for comparison.
The rugged structure likely stems from the finite number of work processes, 
which contribute to the average work $\overline{W}$, and the optimization scheme
itself, identifying only one of the exponentially many close to optimal
protocols \cite{Gingrich_2016}.
Here, the most striking feature is the existence of distinct jumps at the 
beginning and end, present for all sequences at both temperatures.
The jump heights at beginning and end are always different from each other and
are lower at the higher temperature $T=1.0$ compared to $T=0.3$.
Further, the \emph{continuous} and \emph{asymmetric} sequences 
have rather similar optimal protocols,
despite the later one also having partially a primary structure like the
\emph{hairpin} sequence.
Jumps are also observed for simpler systems like brownian-particles
\cite{Schmiedl_2007,Then_2008, Gomez-Marin_2008}, giving further indication that
they are indeed a generic feature of optimized protocols as was already
speculated in Ref.~\cite{Schmiedl_2007}.
Fitting an 5th order polynomial to the protocol to obtain a smoothed version
does not change the qualitative result presented below, i.e. the
work distributions in Fig.~\ref{fig:work:distributions} are effectively
unaltered by using the smoothed protocols.

\begin{figure}
	\includegraphics[width=0.49\linewidth]{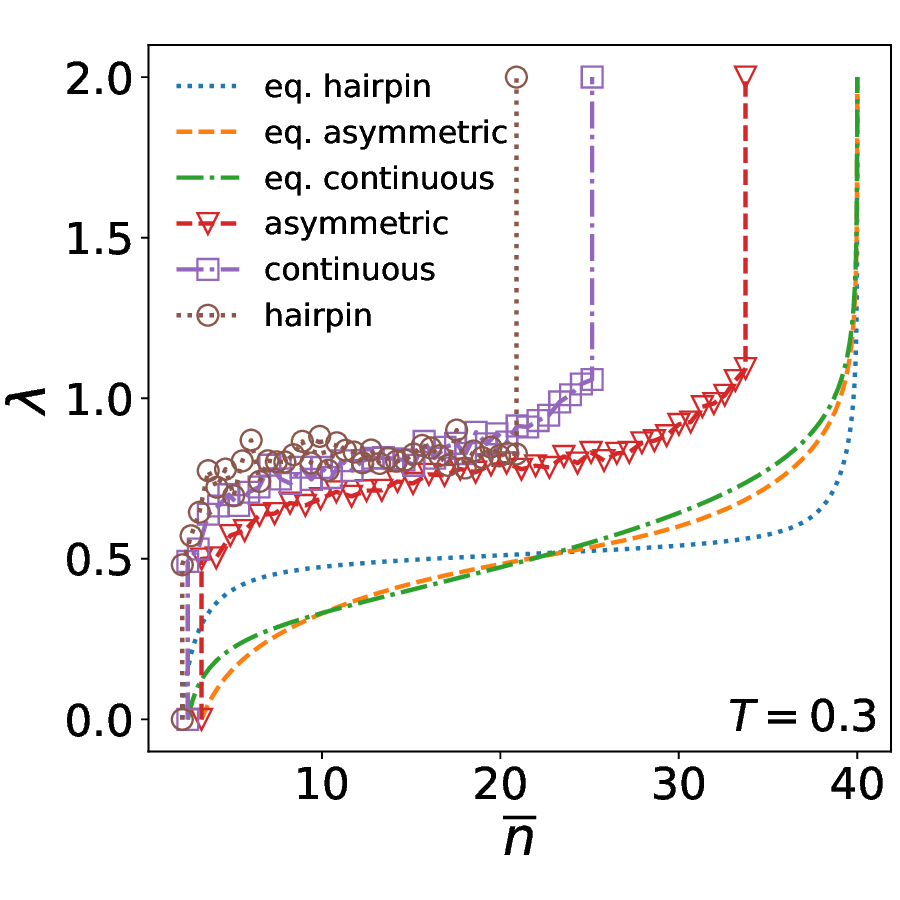}
	\includegraphics[width=0.49\linewidth]{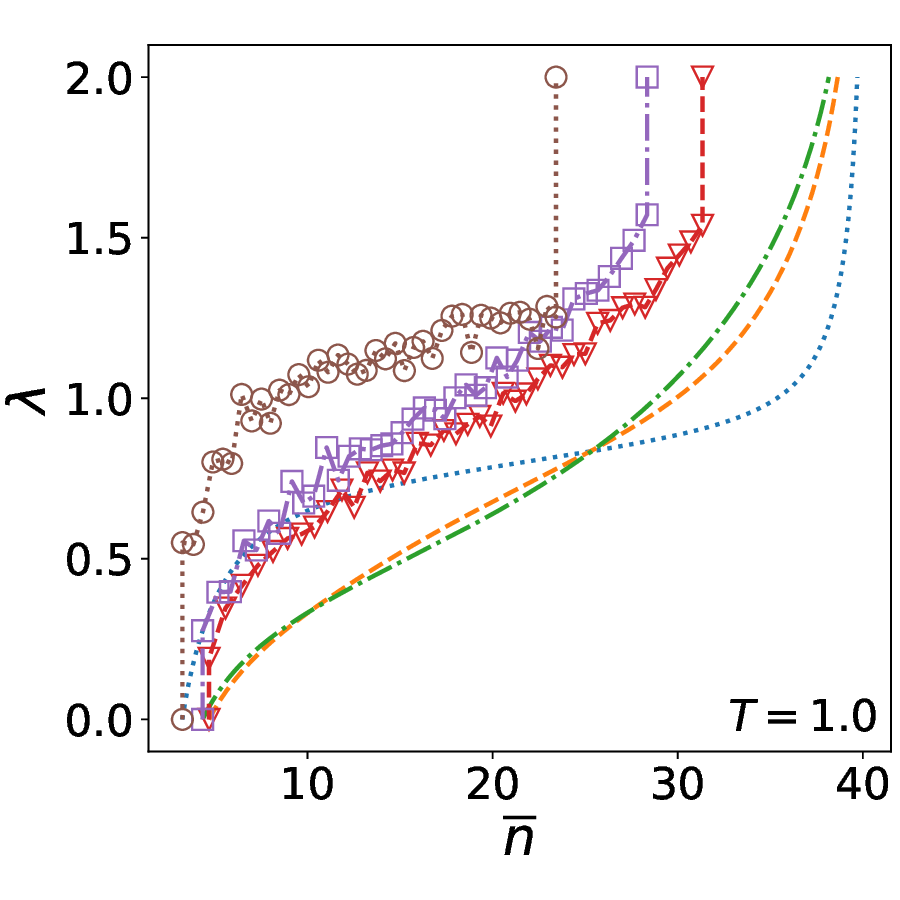}\\
	\includegraphics[width=0.49\linewidth]{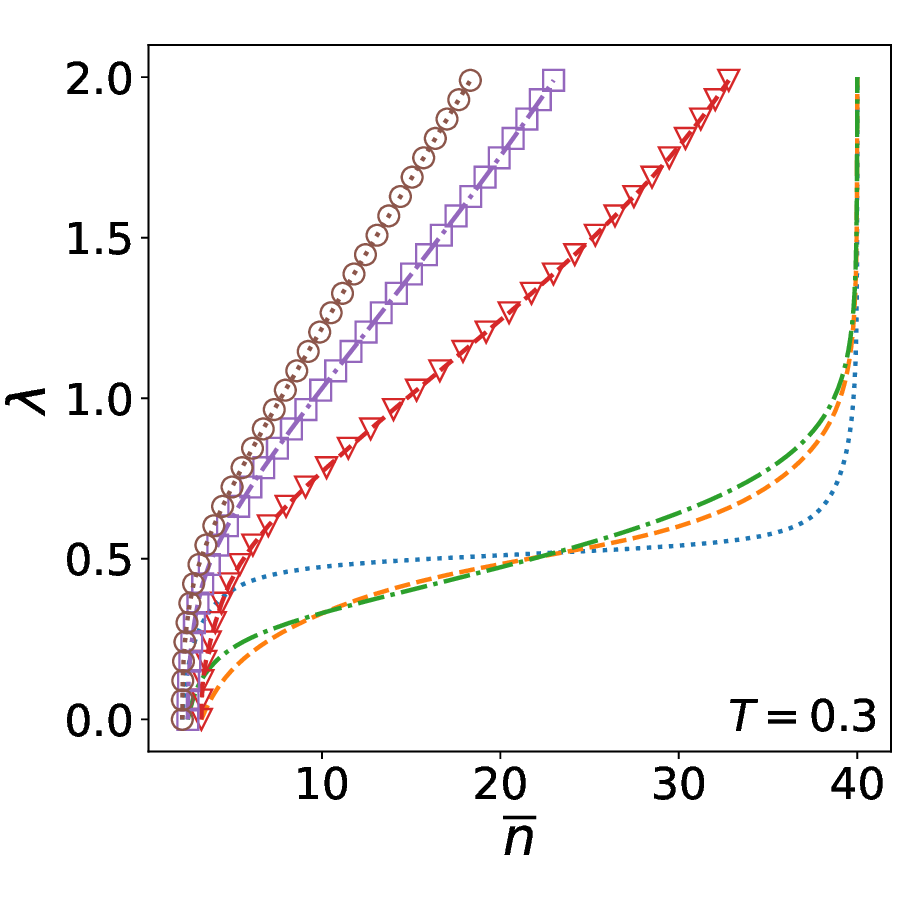}
	\includegraphics[width=0.49\linewidth]{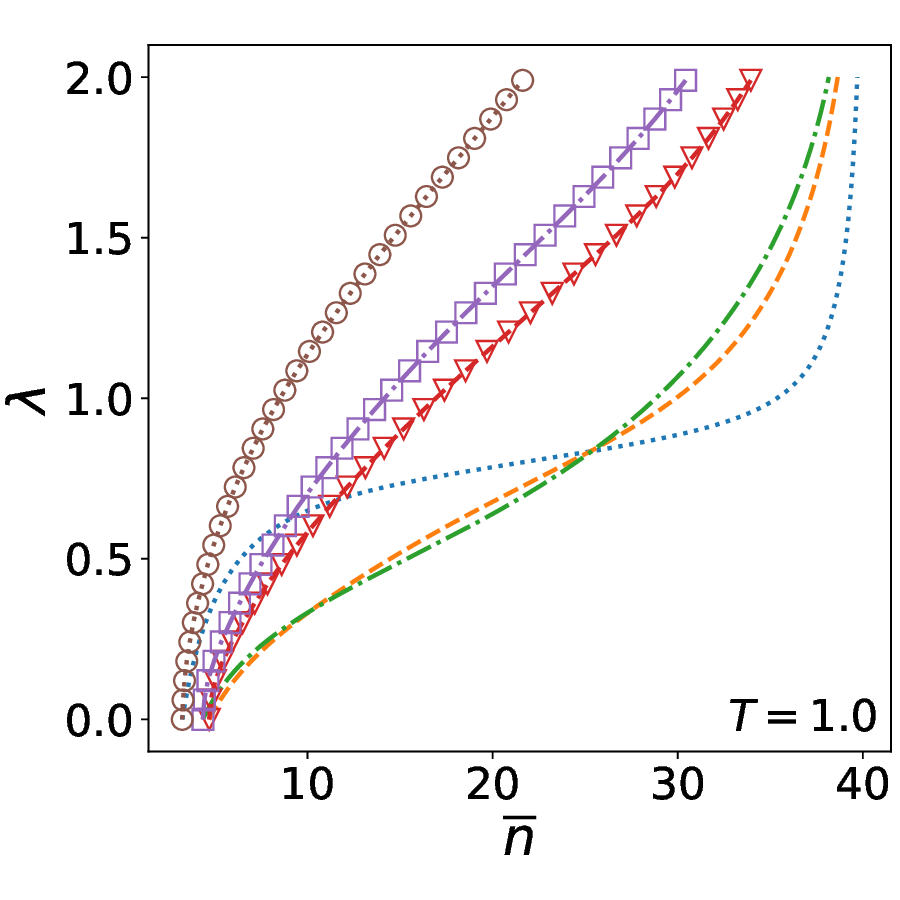}
	\caption{\label{fig:optimized:protocols:mean:free:length:force}
	Protocols $\lambda$ as a function of mean free length $\overline{n}$.
	The equilibrium curves are also shown for reference.
	Top: Optimized protocols.
	The initial jumps are close to the equilibrium unfolding transition point.
	Bottom: Linear protocols for comparison.
	Left: $T=0.3$. Right: $T=1.0$.}
\end{figure}

A more refined analysis is obtained here by recording secondary 
structure trajectories, while the system is subject to the corresponding 
protocol.
This is used to measure the mean free length.
The  force $\lambda$ as a function of mean free length $\overline{n}$,
i.e. $\overline{n}(\lambda)$ with switched axes, is
depicted in Fig.~\ref{fig:optimized:protocols:mean:free:length:force}
together with the equilibrium curves, which are obtained by utilizing the 
partition function.
Here, it becomes apparent how optimality is achieved:
The first jump at the beginning is into the regime of the force unfolding
equilibrium phase transition.
Staying close to the transition point in the flowing, the free length of the
system increases due to interaction with the heat bath, where
energy is taken from.
This allows the second jump at the end to happen with on average higher free
length and therefore resulting in a lower overall work performed.
On the other hand, the vicinity to the critical equilibrium protocol value may 
appear surprising, since one expect the system dynamics to slow down,
which generates more lag to equilibrium and therefore increases dissipation 
\cite{Vaikuntanathan_2009}.
Indeed, opposite behavior, i.e. the avoidance of critical points, has 
been observed for example in the reorientation of spins in an Ising-system \cite{Rotskoff_2015, Gingrich_2016}, where temperature was an additional 
protocol parameter that allows the protocol to circumvent the systems
critical line.
Another major difference is that for the Ising system the process is between
to opposite ordered states, while here the system is driven from folded to
unfolded structures.
Also, the protocol in the present study can not avoid the force unfolding 
transition anyway.
As a result,  varying the protocol only slightly while crossing the critical
point, gives the system more time to equilibrate or staying close to
equilibrium, i.e., reducing the lag between the system current and equilibrium distribution.
For this reason, the optimized protocols also tend to approximate the
equilibrium curves (Fig.~\ref{fig:optimized:protocols:mean:free:length:force} top), especially when compared to the naive linear protocol behaviour
(Fig.~\ref{fig:optimized:protocols:mean:free:length:force} bottom).

\subsection{Trajectories}
\begin{figure}
	\includegraphics[width=0.49\linewidth]{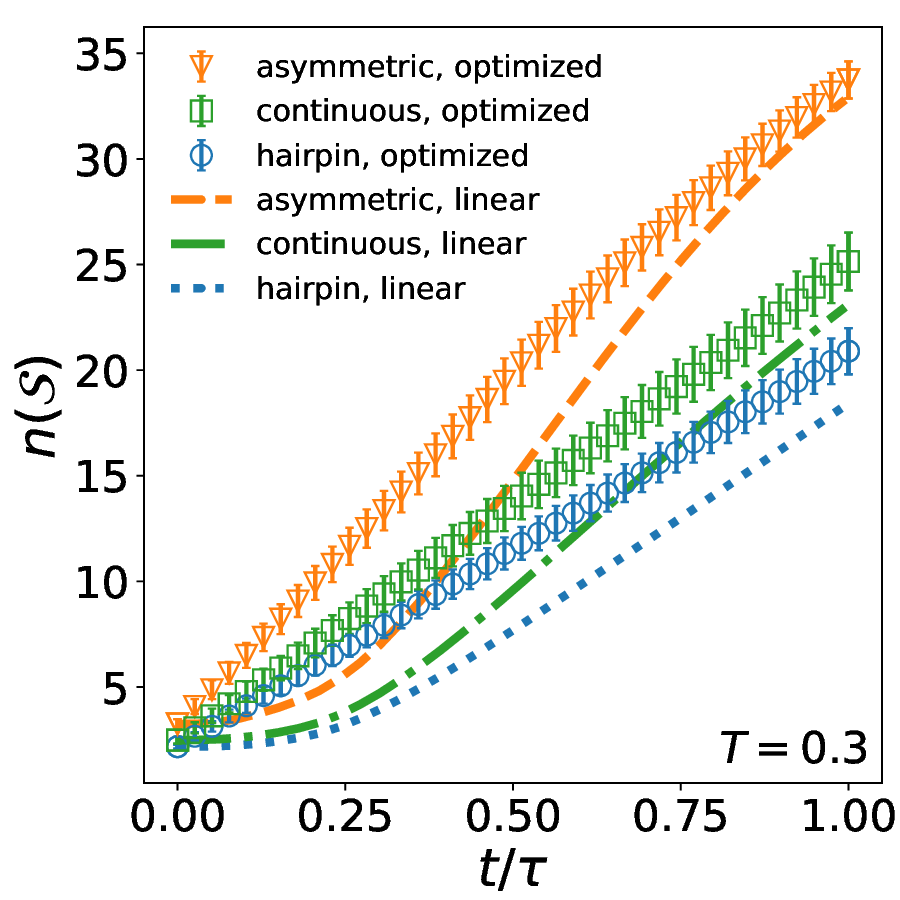}
	\includegraphics[width=0.49\linewidth]{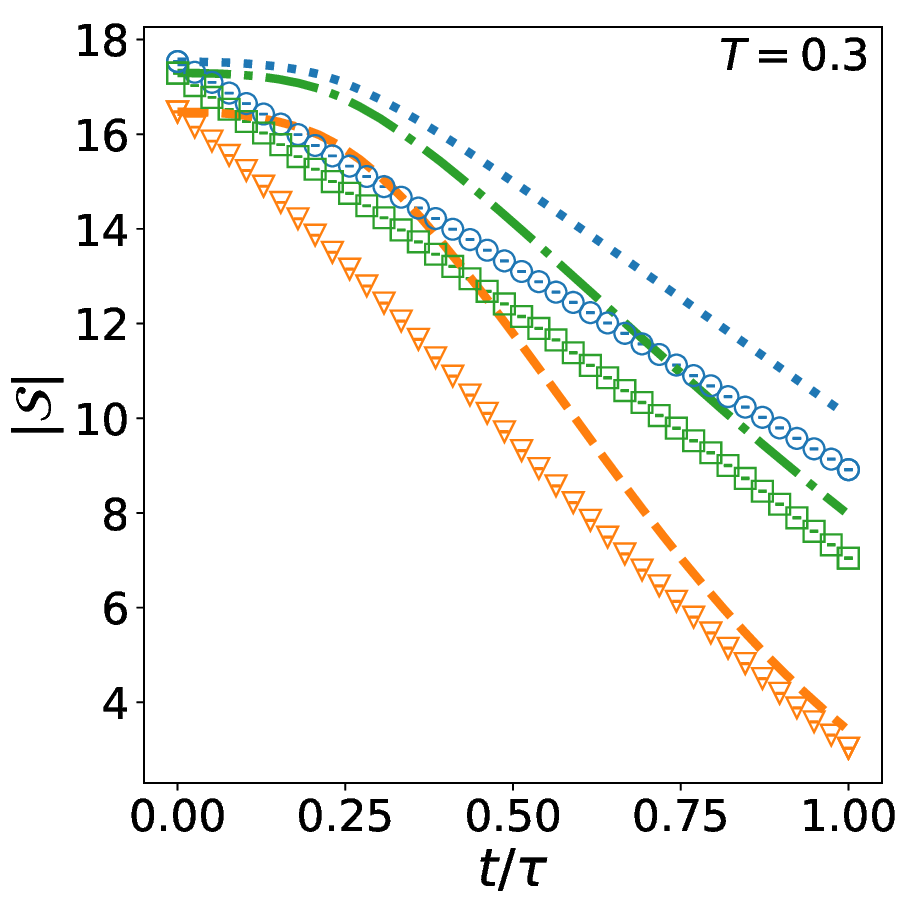}
	\caption{\label{fig:mean:trajectories}Mean system observables as a
	function of time $t$ for $T=0.3$.
	The trajectories for the optimized protocols have an almost linear behavior.
	Left: Mean free length $\overline{n}$. 
	Right: Mean number of paired bases $\overline{\left| \mathcal{S} 
	\right|}$.}
\end{figure}

How does the optimal protocol change the system trajectories compared to a naive
linear protocol?
In Fig.~\ref{fig:mean:trajectories} the mean free length $\overline{n}$ and the
mean number of paired bases $\overline{\left| \mathcal{S} \right|}$ is plotted
as a function of time for both protocols and $T=0.3$.
The $T=1.0$ case looks similar.
Remarkably, the behavior for the optimized protocols is in all cases almost
linear, while the linear protocol results in more sigmoidal trajectories.
The free length for optimized protocol is higher in the beginning than for the
linear protocol, indicating that the unfolding of the secondary structure is
quicker.
In the same train of thought, the number of base pairs decreases for the
optimized protocol faster than for the naive linear protocol.

\subsection{Work Distributions}
\begin{figure}
	\includegraphics[width=\linewidth]{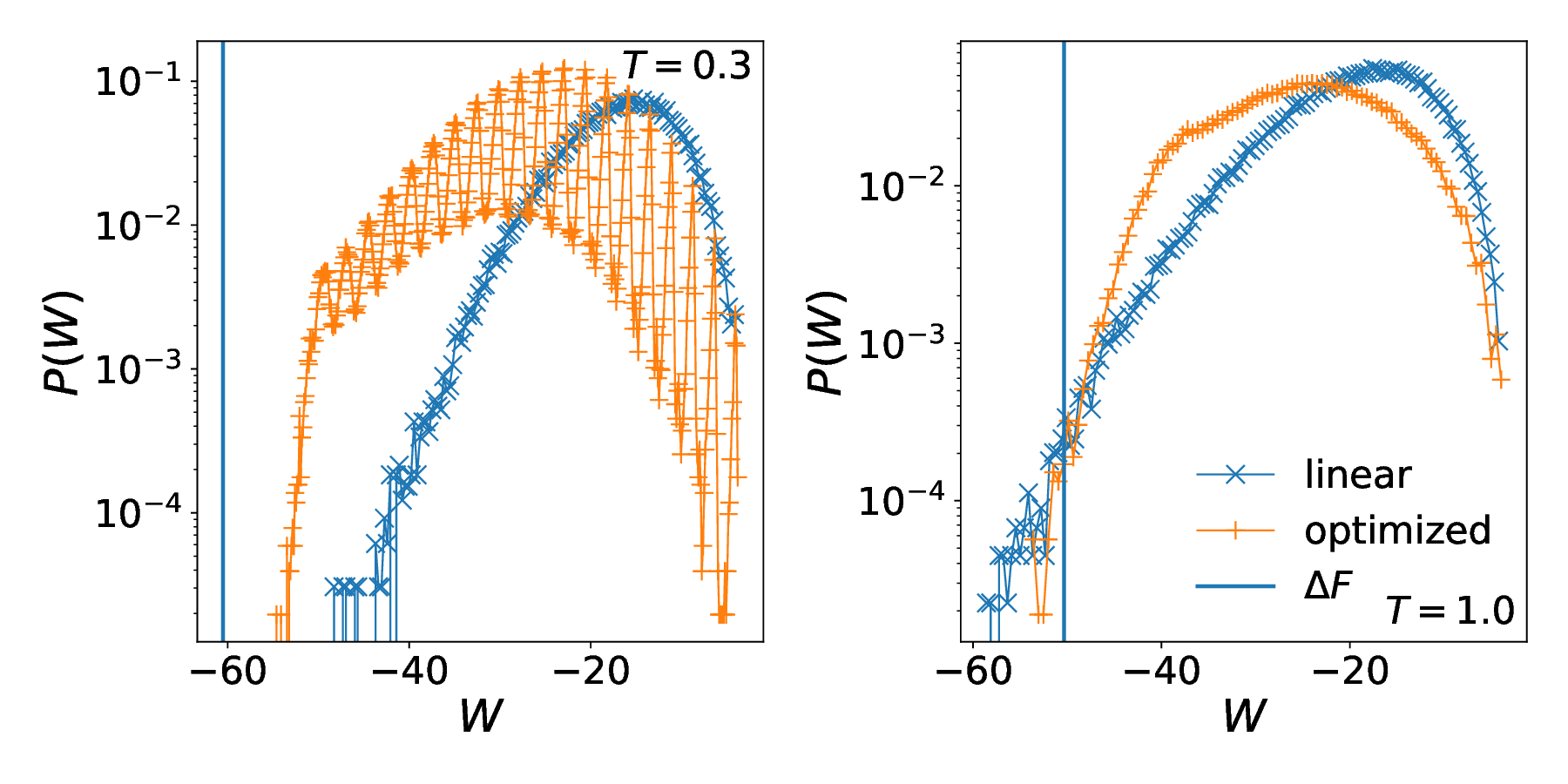}
	\includegraphics[width=\linewidth]{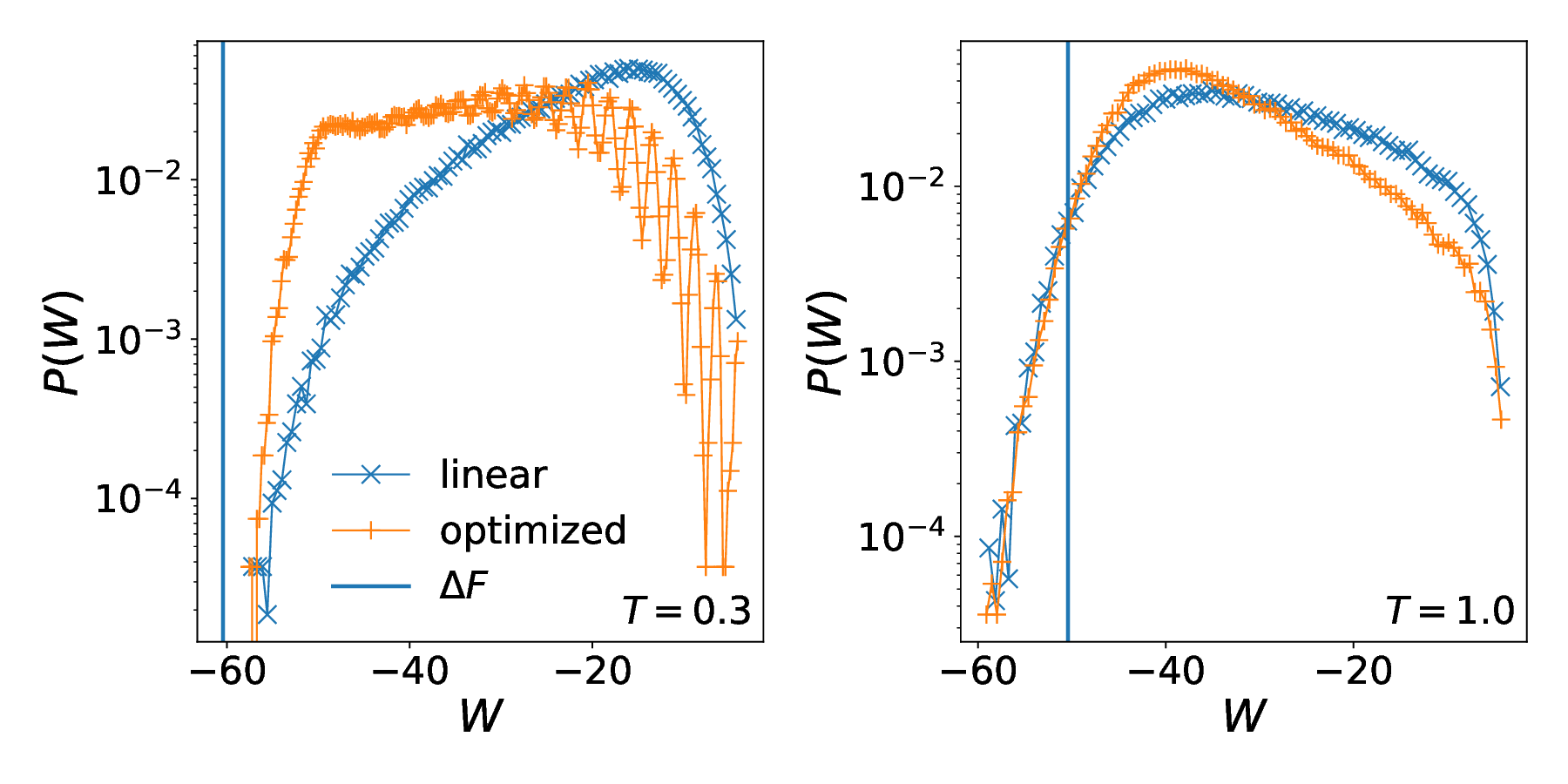}
	\includegraphics[width=\linewidth]{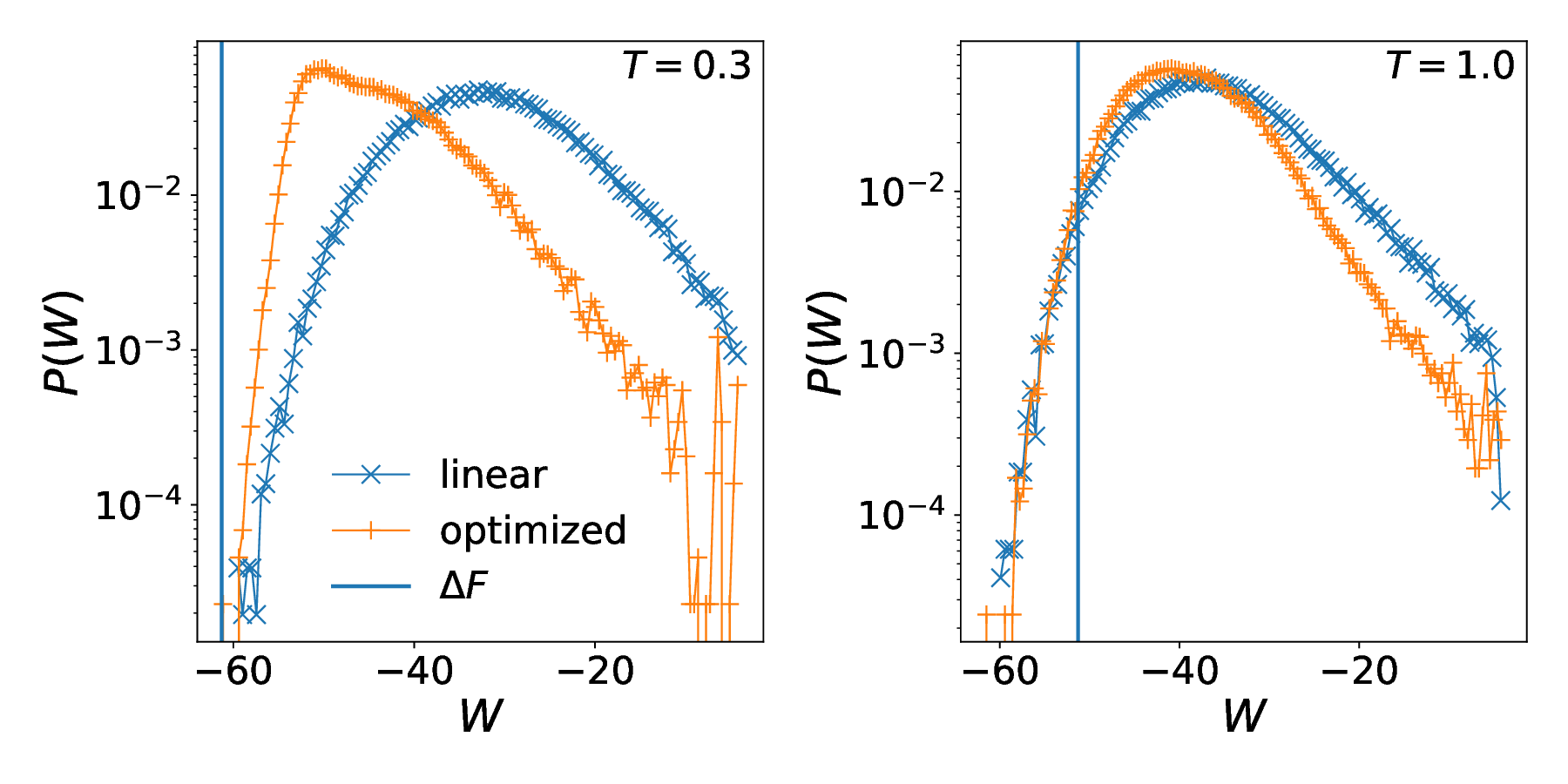}
	\caption{\label{fig:work:distributions}Work distributions for the linear
	and optimized protocols.
	Top: \emph{Hairpin} sequence. 
	Middle: \emph{Continuous} sequence.
	Bottom: \emph{Asymmetric} sequence.
	$10^5$ work processes are used to estimate the histograms, except for the
	optimized protocol at $T=0.3$ for the \emph{hairpin} sequence, where
	$4 \times 10^5$ processes are generated.
	Lines are a guide to the eye only.}
\end{figure}

The resulting work distributions for the linear and optimized protocols are
depicted in Fig.~\ref{fig:work:distributions}.
The general trend is, as expected since the optimized protocols aim at minizing
the work, that the probability is shifted to lower values of $W$.
For $T=1.0$, the peak of the distributions for the optimized protocols, compared
to the linear ones, generally move only slightly but increase in height, while
the tails towards high work values lose statistical weight.
In case of $T=0.3$, not only the peak moves but the distribution shape undergoes
significant change.
The \emph{hairpin} sequence work distribution shows an prominent oscillatory peak
structure as well as the right distribution tail for the continues sequence.
This structure is a result of the discrete free length values, the slow system
dynamics and that the highest contribution to the final work comes from the
initial and final jump, since the protocol changes only slightly in between them.
Work distributions with such erratic behavior have been observed before, e.g.
for the simple model system in \cite{Crooks_1999}, and are therefore not limited
or specific to optimized protocols.\\
Also shown are the exact free energy differences $\Delta F$, directly obtained
from the partition function.
Even a qualitative inspection of the distributions reveals that there is no
substantial improvement compared to the linear protocol case, judging from the
intersection of $P(W)$ with $\Delta F$, which is relevant for its 
estimation when using, e.g., Crooks theorem \cite{Crooks_1999}. 
This underlines the importance of large-deviation algorithms 
\cite{Hartmann_2014,Werner_2021} for such purposes, and that mere 
protocol optimization is not equivalent or even superior.

\begin{figure}
	\includegraphics[width=0.49\linewidth]{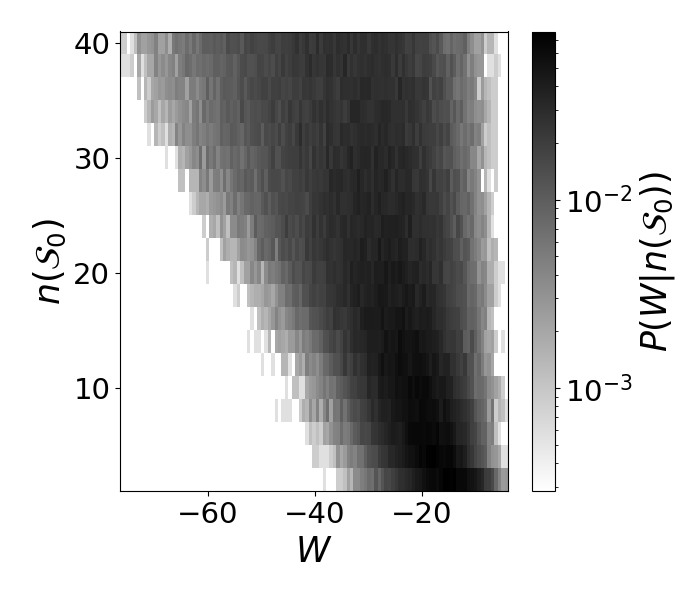}
	\includegraphics[width=0.49\linewidth]{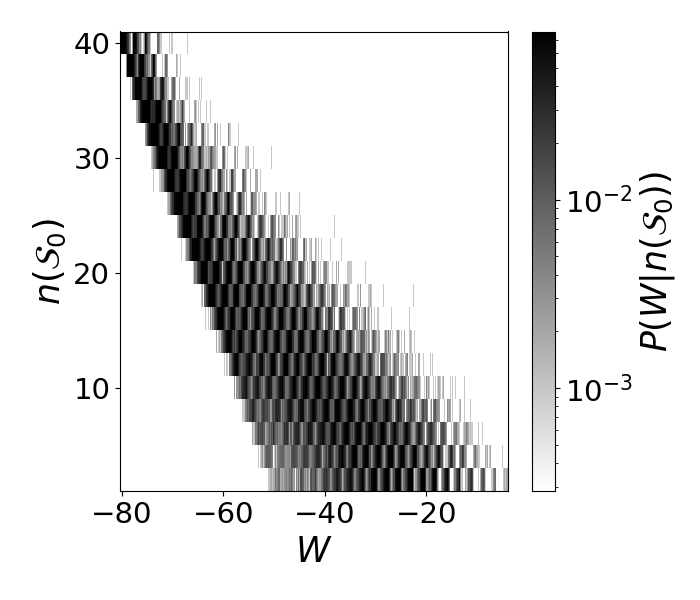}\\
	\includegraphics[width=0.49\linewidth]{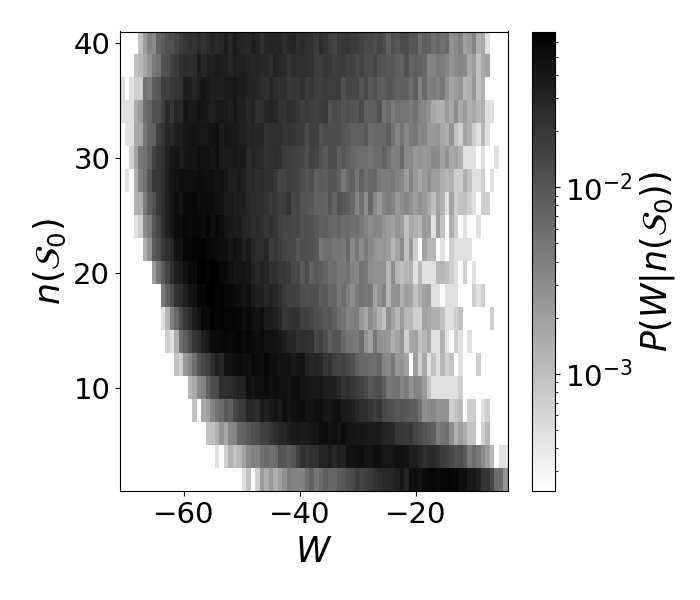}
	\includegraphics[width=0.49\linewidth]{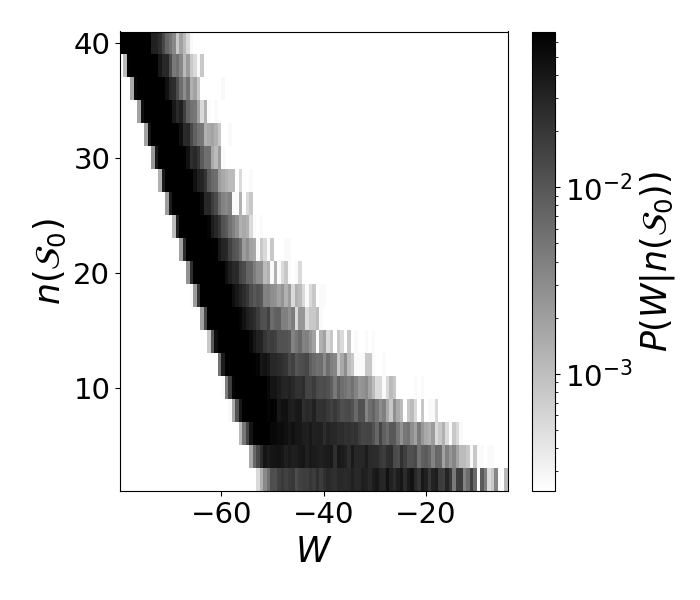}
	\caption{\label{fig:conditional:work:distributions}Shade coded work 
	distributions conditioned to the initial free length $n(\mathcal{S}_0)$ at 
	$T=0.3$.
  A darker shade corresponds to a higher probability.
	For each initial free length 5500 work processes are sampled.
	Left: Linear protocol.
	Right: Optimized protocol.
	Top: \emph{Hairpin} sequence.
	Bottom: \emph{Continuous} sequence.}
\end{figure}

For further investigation of the work distributions, a sampling of the 
initial secondary structures with fixed free length values
$n(\mathcal{S}_0)$ was performed.
The subsequence work processes yielded distributions $P(W|n(\mathcal{S}_0))$
conditioned to said initial free length.
The results are displayed in Fig.~\ref{fig:conditional:work:distributions}).
In comparison to the linear protocol, the optimized protocols especially affect
the work of processes with rare high initial free lengths.
Since the optimized protocol immediately increases the force $\lambda$, 
the resulting structures are less likely to fold towards a more probable state with lower free length during the work process.
This ultimately leads to consistently lower work values for such cases.
For the linear protocol the opposite is true: The low initial force allows the
structure to fold before being unfolded again, resulting in a wide spread of
work values.
For the \emph{continuous} sequence and linear protocol, the weight is more concentrated at smaller values of $W$ as compared to the \emph{hairpin} sequence, most clear for values near $n(\mathcal{S}_0)\approx 26$.
It is due to secondary structures predominantly consisting of multiple small 
hairpins next to each other at the beginning of the work process.
Under an external force, these unfold all individually, i.e. the free length
increases more rapidly than for a single hairpin, because there are more base
pairs that would increase the free length when removed.
This leads on average to an overall lower work compared to the \emph{hairpin}
sequence.

\section{Discussion}
Numerically optimized protocols for simple \emph{hairpin} sequences were
presented for the Higgs-RNA model, which is a model exhibiting many
interacting degrees of freedom.
These optimized protocols show similar distinct jumps as observed before for
systems with only one degree of freedom.
As discussed, the optimized protocols have a connection to the 
unfolding equilibrium phase transition and allow the system to better approximate equilibrium behavior in order to reduce dissipation.
The corresponding work distributions, although shifted to lower
work values, revealed qualitatively that a priori no substantial improvement 
to free-energy estimations can be expected from mere protocol
optimization.
Investigating work distributions conditioned to the initial free length of the
process has shown that especially very unlikely starting values are affected
by optimized protocols.
Due to this last aspect, the employment of a large-deviation algorithm
in combination with optimized protocols, to sample the work distributions in 
the relevant rare event regime, could be a promising perspective.
Also it would be interesting to see whether the results obtained here can be 
found for work processes performed for other complex interacting systems.

\begin{acknowledgments}
This work used the Scientific Compute Cluster at GWDG, the joint data center of
Max Planck Society for the Advancement of Science (MPG) and University of
G\"ottingen.
\end{acknowledgments}

\appendix

\section{Equilibrium Sampling of Secondary Structures with Fixed Free Length}
\label{appendix:equilibrium:sampling:free:length}
For a system given by Eq.~\eqref{eq:total:energy} with force parameter
$\lambda = 0$ and any primary structure $\mathcal{R}$,
the canonical partition function $Z_{i,j}$ can be calculated for all possible
subsequences $r_i, \dots, r_j$ ($i\le j$) via
\begin{equation}\label{eq:partition:function:zero:force}
	Z_{i,j}=Z_{i,j-1} + \sum_{k=i}^{j-s-1} Z_{i,k-1} e^{-\beta e(r_k,r_j)}
	Z_{k+1,j-1}\,,
\end{equation}
using appropriate starting conditions. 
In a similar fashion, the partition
function $Q_{1,j,n}$ at fixed free length $n$ for all subsequences 
$r_1, \dots, r_j$ ($1\le j$) is obtained using 
\begin{equation}\label{eq:partition:function:free:length}
Q_{1,j,n}=Q_{1,j-1,n-1}+ \sum_{k=n-1}^{j-s-1} Q_{1,k-1,n-2} 
e^{-\beta e(r_k,r_j)}Z_{k+1,j-1}\,,
\end{equation}
again with compatible starting values.
All of this can be done using \textit{dynamical programming} in $O(L^3)$ runtime
\cite{Nussinov_1978}.
By diving Eq.~\eqref{eq:partition:function:free:length} by $Q_{1,j,n}$,
the individual terms can be identified as pairing probabilities.
First, the probability of base $j$ being unpaired:
\begin{equation}\label{eq:pairing:probabilities:unpaired}
	p^u_{j,n} = \frac{Q_{1,j-1,n-1}}{Q_{1,j,n}}.
\end{equation}
The remaining terms are the probabilities that base $j$ is paired to any base
$k$ with $n-1 \le k \le j-s-1$:
\begin{equation}\label{eq:pairing:probabilities:paired}
	p^p_{j,k,n} = \frac{Q_{1,k-1,n-2} e^{-\beta e(r_k,r_j)}Z_{k+1,j-1}}
	{Q_{1,j,n}}.
\end{equation}
The sampling is then done via the recursive scheme described in 
\cite{Werner_2021} and roughly goes as follows:
For a given subsequence from base $i$ to $j$ with so far no pair, 
beginning with $i=1$ and $j=L$, base $j$ is randomly paired to another base $k$ in the subsequence using the correspondig pairing probabilities.
If base $j$ remains unpaired, the scheme is continued by setting $j'=j-1$.
Should base $j$ be paired to base $k$, the recursive procedure is independently 
continued for the two created subsequence from $i'=i$ to $j'=k-1$ and $i'=k+1$ 
to $j'=j-1$.

\bibliography{references}

\end{document}